\documentclass[preprint,10pt]{elsarticle}

\makeatletter
\def\ps@pprintTitle{%
\let\@oddhead\@empty
\let\@evenhead\@empty
\def\@oddfoot{\centerline{\thepage}}%
\let\@evenfoot\@oddfoot}
\makeatother

\journal{Physica D}
\usepackage{graphicx,bbm,setspace,amssymb,amsfonts,amsmath,mathtools,mathrsfs,stmaryrd,amsthm,bm,etoolbox,comment,hyperref,url,caption,subcaption,color,enumitem,algorithm,booktabs,microtype,nicefrac,lingmacros,nccmath,tree-dvips,tikz-cd}

\usepackage[colorinlistoftodos]{todonotes}
\usepackage[utf8]{inputenc} % allow utf-8 input
\usepackage[T1]{fontenc}    % use 8-bit T1 fonts

\usepackage[noend]{algpseudocode}

\newtheorem{theorem}{Theorem}[section]

\newtheorem{thm-defn}[theorem]{Theorem/Definition}

\theoremstyle{definition}

\theoremstyle{remark}
\newtheorem{remark}[theorem]{Remark}

\newcommand{\ignore}[1]{}{}

\begin{document}

\begin{frontmatter}

\title{Trends in COVID-19 prevalence and mortality: a year in review}
   
\author[label1]{Nick James} \ead{nick.james@unimelb.edu.au}
\author[label2]{Max Menzies} 
\address[label1]{School of Mathematics and Statistics, University of Melbourne, Victoria, Australia}
\address[label2]{Yau Mathematical Sciences Center, Tsinghua University, Beijing, China}

\begin{abstract}
This paper introduces new methods to study the changing dynamics of COVID-19 cases and deaths among the 50 worst-affected countries throughout 2020. First, we analyse the trajectories and turning points of rolling mortality rates to understand at which times the disease was most lethal. We demonstrate five characteristic classes of mortality rate trajectories and determine structural similarity in mortality trends over time. Next, we introduce a class of \emph{virulence matrices} to study the evolution of COVID-19 cases and deaths on a global scale. Finally, we introduce \emph{three-way inconsistency analysis} to determine anomalous countries with respect to three attributes: countries' COVID-19 cases, deaths and human development indices. We demonstrate the most anomalous countries across these three measures are Pakistan, the United States and the United Arab Emirates.

\end{abstract}


\end{frontmatter}

%------------BODY OF PAPER ------------------
\section{Introduction}
\label{sec:Introduction}

%1. Summary of 2020 and Covid - fine as is
2020 will be remembered as the year that the world first battled the COVID-19 pandemic. Almost 2 million people lost their lives, substantial restrictions on population movement and activities were imposed, and almost every country experienced an economic recession. During that year, treatments improved substantially \cite{Remdesivir,Bloch2020,toczilizumab,Cao2020} and several vaccines were produced by the end of the year \cite{Polack2020,Walsh2020}. However, the disease remains highly prevalent around the world as of the start of 2021, and measures to contain and reduce its transmission remain highly relevant for the reduction of casualties as well as economic and other social consequences \cite{Priesemann2021,Momtazmanesh2020}.

%2. Different government responses - fine as is
Throughout the year, government responses to the pandemic varied substantially, both over time and between countries. Early government responses included banning travel \cite{bbccloseborders_2020}, the implementation of testing and contact tracing programs \cite{Koreaguardian_2020}, and lockdowns. Due to the economic consequences of lockdowns, many countries implemented them too late \cite{italynyt2020, Scally2020} and lifted restrictions before cases had sufficiently reduced \cite{wapo_allreopen}. Such disparate responses to the virus led to great variability in case and death counts, creating \emph{different waves} of the outbreak across many countries. Such later waves often exhibited higher case and death counts than the first \cite{atlantic_secondsurge, james2020covidusa}.

%3. NEW PARA: What previous Covid research has done.
The response of the scientific community to COVID-19 has also been varied and multifaceted, producing research from many perspectives and disciplines. 
%to respond to the unprecedented impact and ongoing threat of the pandemic. 
In addition to the aforementioned medical research \cite{Remdesivir,Bloch2020,toczilizumab,Cao2020,Polack2020,Walsh2020}, mathematical approaches to model and analyse the virus and its impact have been broad. First, models based on existing statistical techniques, such as the Susceptible–Infected–Recovered (SIR) model and the basic reproductive ratio $R_0$, have been proposed and systematically collated by researchers \cite{Wynants2020,ModellingEstrada2020}. These have been used for various purposes, including diagnosis and prognosis of COVID-19 patients, efficacy of medications, and vaccine development. Next, nonlinear dynamics researchers have proposed several sophisticated extensions to the classical predictive SIR model, including analytic techniques to find explicit solutions \cite{SIRBarlow2020, SIRWeinstein2020}, modifications to the SIR model with additional variables \cite{SIRNg2020,SIRVyasarayani2020,SIRCadoni2020,SIRNeves2020,SIRComunian2020}, incorporation of Hamiltonian dynamics \cite{SIRBallesteros2020} or network models \cite{SIRLiu2021} and a closer analysis of uncertainty in the SIR approach \cite{Gatto2021}. Other mathematical approaches to prediction and analysis include power-law models \cite{Manchein2020,Blasius2020,Beare2020}, distance analysis \cite{daSilva2021,James2020_nsm}, network models \cite{Shang2020, Karaivanov2020,Ge2020,Xue2020}, analyses of the dynamics of transmission and contact \cite{Saldaa2020,Danchin2021}, forecasting models \cite{Perc2020}, Bayesian methods \cite{Manevski2020}, clustering \cite{Machado2020,jamescovideu} and many others \cite{Ngonghala2020,Cavataio2021,Nraigh2020,Glass2020}.

%4. Intro to motivation/originality/contributions/synergy
We have a different motivation and approach relative to the aforementioned work. Rather than performing predictions on an individual country basis or comparing parameters among different countries (such as $R_0$ or power-law exponents), we seek to reveal structural similarity in COVID-19 case, death and mortality time series across many countries of the world. Rather than predicting the future, which is always challenging due to unpredictable changes in government policy, we aim to be descriptive, revealing similarity and anomalous countries in outcomes. Indeed, close analysis of the case, death and mortality dynamics on a country-by-country basis is necessary to inform governments of the most successful strategies for reducing transmission of cases and progression to deaths. Identifying structural similarities between countries' trajectories can support conclusions that certain government responses will likely result in better or worse outcomes. Moreover, identifying anomalous countries can provide insights on which responses to the pandemic were exceptionally good or poor. For this purpose, we present three sections, each of which contributes a new mathematical method for analysing the world's COVID-19 cases, deaths and mortality, or any multivariate time series more generally.

%5. Structure paragraph - ADD IN ORIGINALITY/CONTRIBUTIONS
This paper is therefore structured as follows. In Section \ref{sec:deathrates}, we analyse the trajectories of mortality rates on a country-by-country basis. In particular, we build upon a recently introduced algorithmic framework to identify the \emph{turning points} of the mortality trajectories, which reveal when the disease was most and least lethal (with respect to the progression from cases to deaths). We then use a new semi-metric between finite sets to assign countries into classes of mortality trajectories. We believe this is the first work to classify different mortality trajectories among countries, rather than a more traditional comparison of overall mortality rates without considering its changing dynamics over time. In Section \ref{sec:virulencematrices}, we analyse the eigenspectra of \emph{virulence matrices} as a new means of understanding trends in the worldwide prevalence and mortality of COVID-19. 
This reveals periods in which COVID-19 was most severe and most heterogeneous between countries.
%\textcolor{blue}{This is the first paper to analyse the world's case, death, and mortality numbers on a country-by-country basis in conjunction in this way, to reveal periods in which COVID-19 was most severe and most homogeneous between countries.}  NOW IN CONCLUSION
In Section \ref{sec:3wayCCA}, we compare countries' case and death counts with their \emph{human development index (HDI)} and use a new method to identify the most anomalous countries between these attributes. 
%\textcolor{blue}{We believe that this is the first paper to quantitatively combine these features and identify respective anomalies.} NOW IN CONCLUSION
In Section \ref{sec:conclusion}, we discuss our findings from the aforementioned analyses regarding COVID-19 trends throughout the year 2020. We conclude in Section \ref{sec:finalconclusion}.

%--------------------SECTION II ----------------
\section{Mortality rate analysis}
\label{sec:deathrates}

%Para - intro to the TS we're studying
In this section, we study the dynamics of the COVID-19 mortality rate among $n=50$ countries. Our data spans 01/01/2020 to 31/12/2020, a period of $T=366$ days. We choose the countries with the 50 greatest total case counts of COVID-19 as of 31/12/2020, order these by alphabetical order, and index them $i=1,...,n$. Let $x_i(t), y_i(t) \in \mathbb{R}$ be the multivariate time series of new daily cases and deaths, respectively, for $i=1,..., n$ and $t=1,...,T$. Throughout this paper, the subscript $i$ pertains to the $i$th country, ordered alphabetically, while evaluating a function at $t$ gives its value at the $t$th day of the year. For a given country, let $r_i(t)$ be its 30-day rolling mortality rate, defined by
\begin{align}
    r_i(t)=\dfrac{\sum_{s=t-29}^{t}  y_i(s)}{\sum_{s=t-29}^{t} x_i(s)}, t=30,..., T,
\end{align}
or zero if no cases have been observed over the last 30 days. 
%Varying this across countries 
This gives a multivariate time series $r_i(t)$, for $i=1,..., n$ and $t=30,...,T$. The data point at time $t$ describes the rolling mortality rate over the prior 30 days. 

\subsection{Methodology}

%Para - general similarity and HC
The aim of this section is to study these mortality trends on a country-by-country basis and identify structural similarity across different countries. For this purpose, we use two (semi)-metrics between the mortality rate time series and apply \emph{hierarchical clustering} \cite{Ward1963,Szekely2005} to these measures. Hierarchical clustering has been used in several epidemiological applications, including inflammatory diseases \cite{Madore2007}, airborne diseases \cite{Kretzschmar2009}, Alzheimer's disease \cite{Alashwal2019},  Ebola \cite{Muradi2015}, SARS \cite{Rizzi2010}, and COVID-19 \cite{Machado2020}.

%Para - introduce waves and turning points
These mortality rates $r_i(t)$ exhibit highly undulating behaviour, moving between clear peaks and troughs (turning points). Our first semi-metric measures distance between algorithmically-identified turning points as a proxy for each time series' behaviour. We modify an existing algorithmic framework for this purpose. First, we apply a \emph{Savitzky-Golay filter} to produce a collection of smoothed time series $\hat{r}_i(t)$, $t=30,...,T$ and $i=1,...,n$. A beneficial effect of this smoothing is to ameliorate some of the noise present in the COVID-19 case count data. In addition to the smoothing procedure, our computation includes a rolling mortality rate that further reduces the effect of perturbations in the data's underlying signal. We choose a 30-day rolling mortality rate for two reasons: first, this window length provides a compromise between denoising the data and not over-smoothing; %such that true variability in the signal is lost. 
second, 30 days provides a good luck at the mortality rate behaviour of the last month of data. Next, we follow \cite{james2020covidusa} and apply a two-step algorithm where we select and then refine a set of turning points. We assign each smoothed mortality rate time series a non-empty set $P_i$ and $T_i$ of local maxima (peaks) and local minima (troughs). To better suit our specific application, we modify the second step of this algorithm, in which the turning point list is refined. Full details are included in \ref{appendix:turningpoint}, including a discussion of the procedure's robustness against noisy data. We display 12 countries' mortality rate time series and annotate their turning points in Figure \ref{fig:DeathRateTrajectories}.

To quantify distance between time series' turning points, we modify the semi-metric of \cite{James2020_nsm} (with $p=1$). Given two non-empty finite sets $S_1,S_2 \subset \{1,2,...,T\}$, this is defined as 
\begin{align}
\label{eq:MJ}
    D({S_1},{S_2}) = \frac{1}{2T} \left(\frac{\sum_{b\in S_2} d(b,S_1)}{\#S_2} + \frac{\sum_{a \in {S_1}} d(a,S_2)}{\#S_1} \right),
\end{align}
where $d(b,S_1)$ is the minimal distance from $b \in S_2$ to the set $S_1$, and $\# S_1$ is the cardinality of $S_1$, and analogously for $S_2$. By the choice of normalisation, this is always bounded between 0 and 1. To more appropriately separate different behaviours among mortality trends, we modify this semi-metric by including a regularisation term. This treatment is inspired by various regularisation penalties in the statistical literature \cite{Zou2005}. We construct our semi-metric as follows:
\begin{align}
\label{eq:newMJ}
    D'({S_1},{S_2}) = D(S_1,S_2) + \beta|\#S_1 - \#S_2|,
\end{align}
where $0<\beta\leq1$ is a constant.  The resulting values $D'(S_1,S_2)$ are symmetric, non-negative, and zero if and only if $S_1=S_2$. Then, we define the $n \times n$ matrix $D^{TP}$ between turning point sets by
\begin{align}
D_{ij}^{TP} = D'(P_i,P_j) + D'(T_i,T_j).
\end{align}
Here, $P_i$ denotes the set of peaks for the $i$th country's mortality series, while the subscript $ij$ gives a distance between countries $i$ and $j$, ordered alphabetically. In Figure \ref{fig:DeathRateDendrogramMJ1}, we perform hierarchical clustering on $D^{TP}$ with a range of values of $\beta$. These distances do not capture the absolute values of the mortality rate time series; they only distinguish between their undulating behaviour, reflected in their sets of turning points. To round out our analysis, we include another metric, an $L^1$ norm that does account for difference in the absolute values of mortality. We define another matrix by 
\begin{align}
D_{ij}^{1}= \sum_{t=30}^T |r_i(t)-r_j(t)|,
\end{align}
and perform hierarchical clustering on $D^1$ in Figure \ref{fig:DeathRateDendrogramTrajectory}. Again, the subscript $ij$ refers to a distance computed between countries $i$ and $j$, ordered alphabetically.

\subsection{Results}
\begin{figure}
    \centering
    \begin{subfigure}[b]{0.327\textwidth}
        \includegraphics[width=\textwidth]{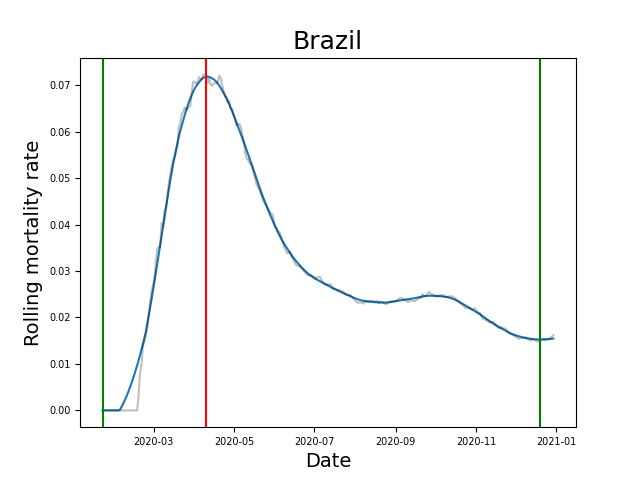}
        \caption{}
        \label{fig:DeathRateBrazil}
    \end{subfigure}
    \begin{subfigure}[b]{0.327\textwidth}
        \includegraphics[width=\textwidth]{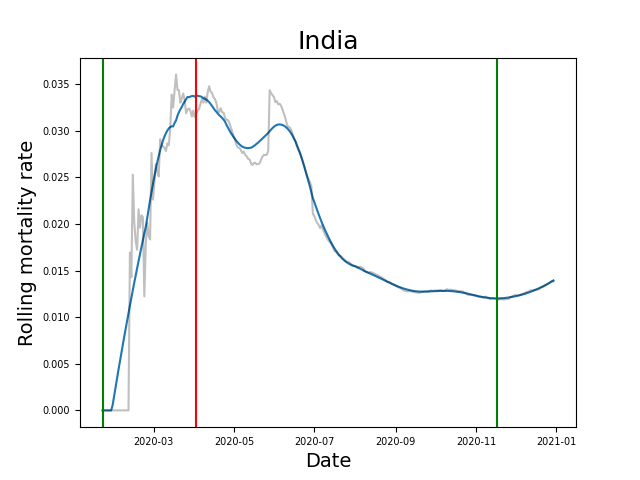}
        \caption{}
        \label{fig:DeathRateIndia}
    \end{subfigure}
        \begin{subfigure}[b]{0.327\textwidth}
        \includegraphics[width=\textwidth]{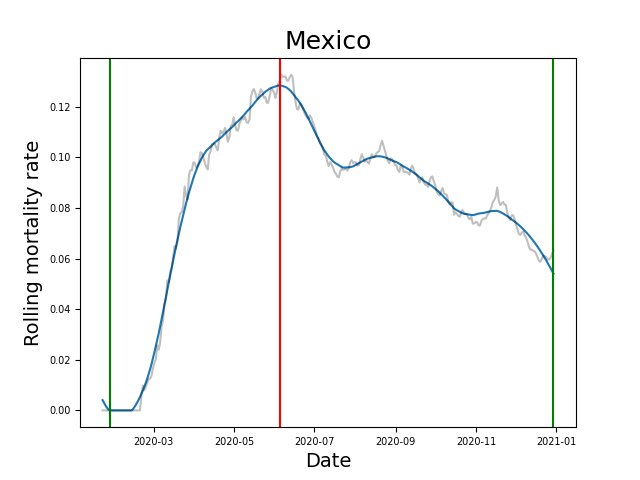}
        \caption{}
        \label{fig:DeathRateMexico}
    \end{subfigure}
        \begin{subfigure}[b]{0.327\textwidth}
        \includegraphics[width=\textwidth]{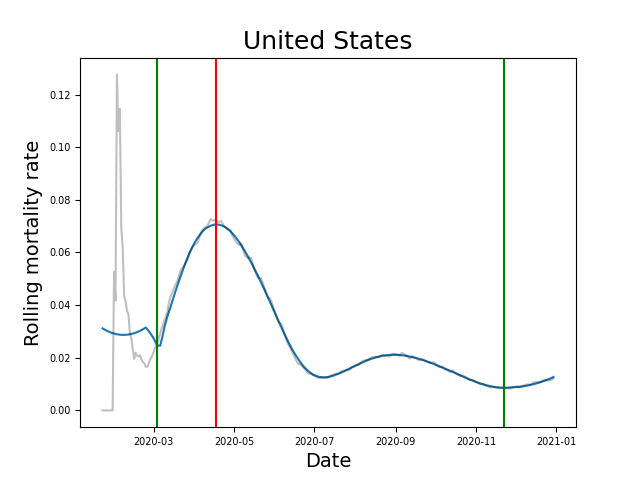}
        \caption{}
        \label{fig:DeathRateUS}
    \end{subfigure}    
    \begin{subfigure}[b]{0.327\textwidth}
        \includegraphics[width=\textwidth]{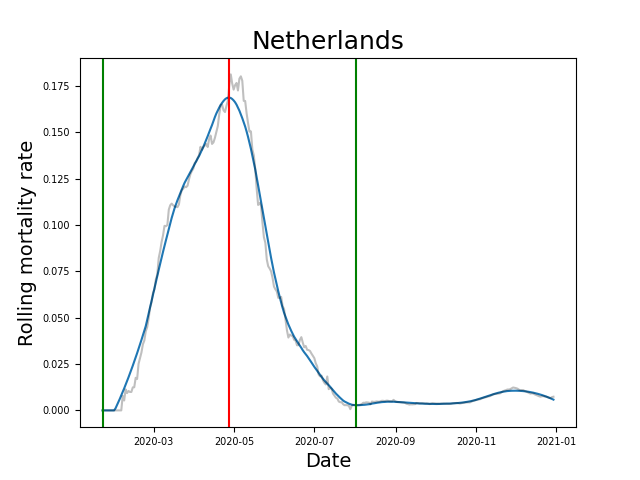}
        \caption{}
        \label{fig:DeathRateNetherlands}
    \end{subfigure}    
    \begin{subfigure}[b]{0.327\textwidth}
        \includegraphics[width=\textwidth]{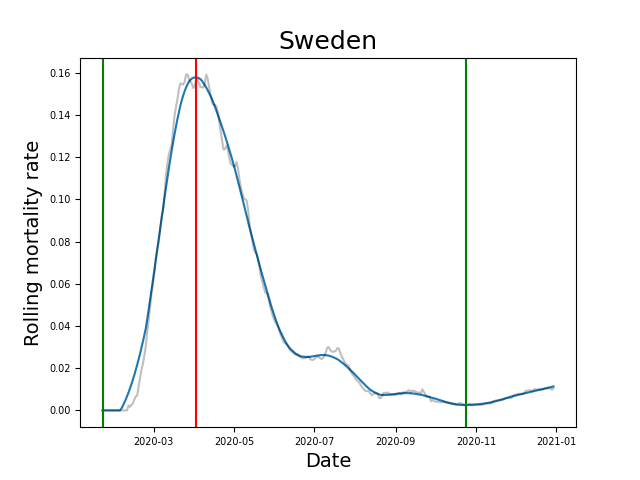}
        \caption{}
        \label{fig:DeathRateSweden}
    \end{subfigure}         
    % \begin{subfigure}[b]{0.327\textwidth}
    %     \includegraphics[width=\textwidth]{Death_rate_France_smoothed_.png}
    %     \caption{}
    %     \label{fig:DeathRateFrance}
    % \end{subfigure}
    \begin{subfigure}[b]{0.327\textwidth}
        \includegraphics[width=\textwidth]{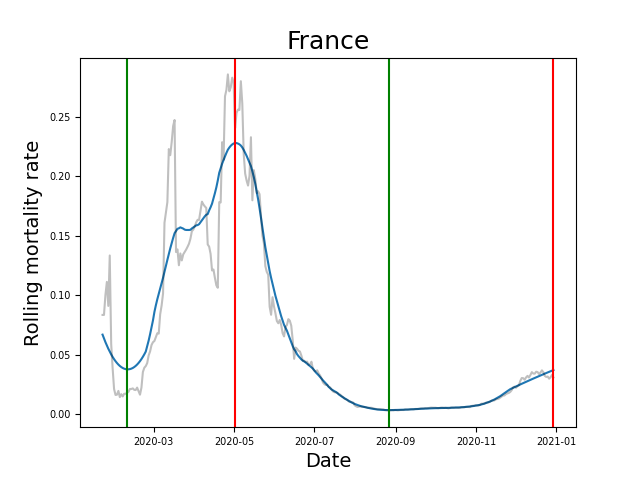}
        \caption{}
        \label{fig:DeathRateFrance}
    \end{subfigure}
    \begin{subfigure}[b]{0.327\textwidth}
        \includegraphics[width=\textwidth]{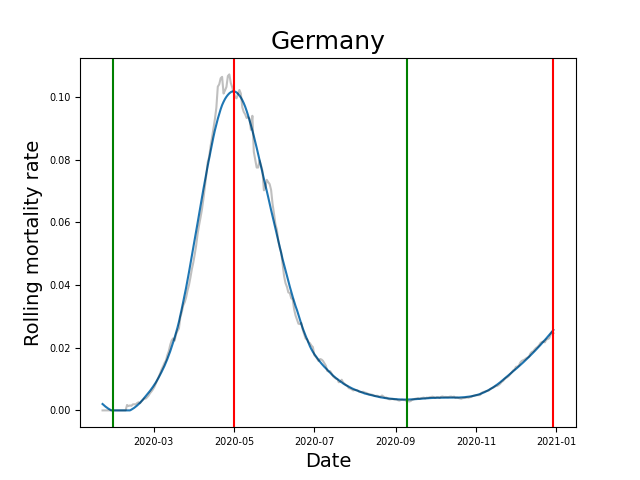}
        \caption{}
        \label{fig:DeathRateGermany}
    \end{subfigure}
        \begin{subfigure}[b]{0.327\textwidth}
        \includegraphics[width=\textwidth]{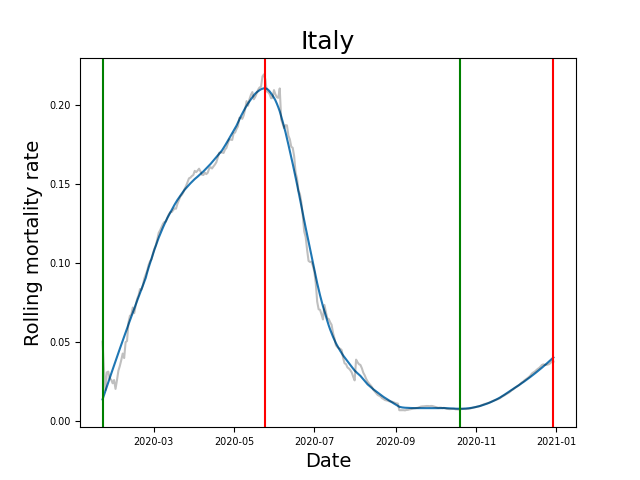}
        \caption{}
        \label{fig:DeathRateItaly}
    \end{subfigure}    
         \begin{subfigure}[b]{0.327\textwidth}
        \includegraphics[width=\textwidth]{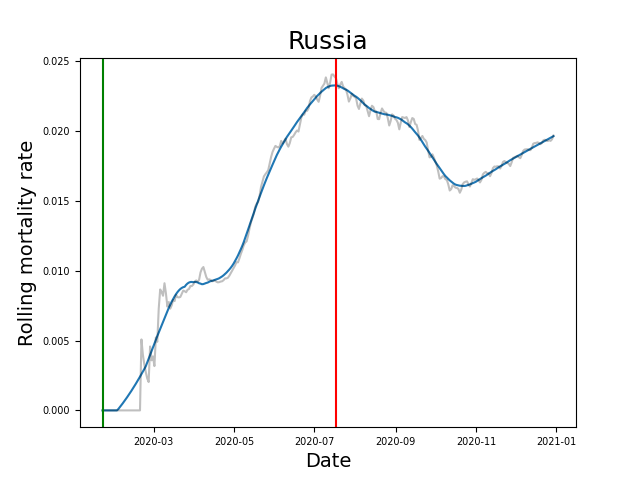}
        \caption{}
        \label{fig:DeathRateRussia}
    \end{subfigure}
    \begin{subfigure}[b]{0.327\textwidth}
        \includegraphics[width=\textwidth]{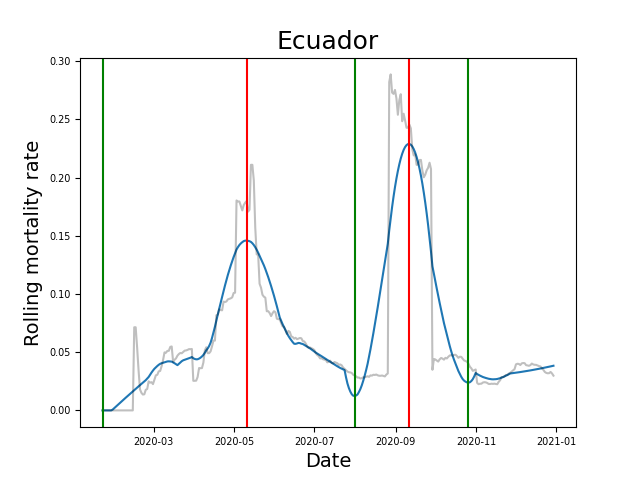}
        \caption{}
        \label{fig:DeathRateEcuador}
    \end{subfigure}
    \begin{subfigure}[b]{0.327\textwidth}
        \includegraphics[width=\textwidth]{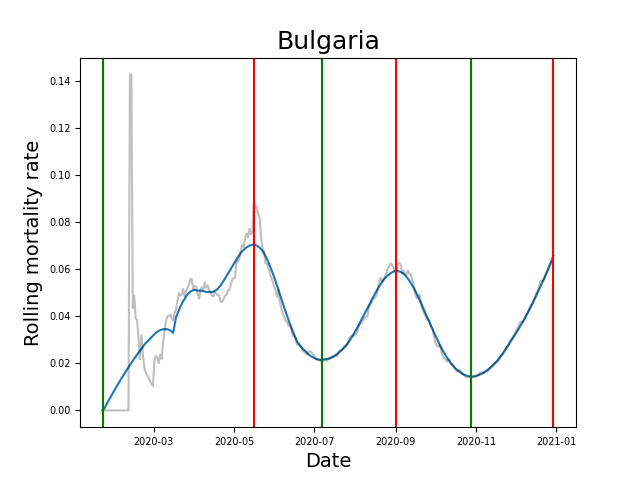}
        \caption{}
        \label{fig:DeathRateBulgaria}
    \end{subfigure}
    \caption{Smoothed mortality rate time series and identified turning points for various countries: (a) Brazil (b) India (c) Mexico (d) the US (e) the Netherlands (f) Sweden (g) France (h) Germany (i) Italy (j) Russia (k) Ecuador (l) Bulgaria. Green and red vertical lines denote algorithmically detected troughs and peaks, respectively. The rolling mortality rate at a given time calculates the mortality over the previous 30 days. The aforementioned countries represent at least one member of every characteristic class of trajectories.}
    \label{fig:DeathRateTrajectories}
\end{figure}

\begin{figure*}
    \centering
    \begin{subfigure}[b]{\textwidth}
        \includegraphics[width=0.905\textwidth]{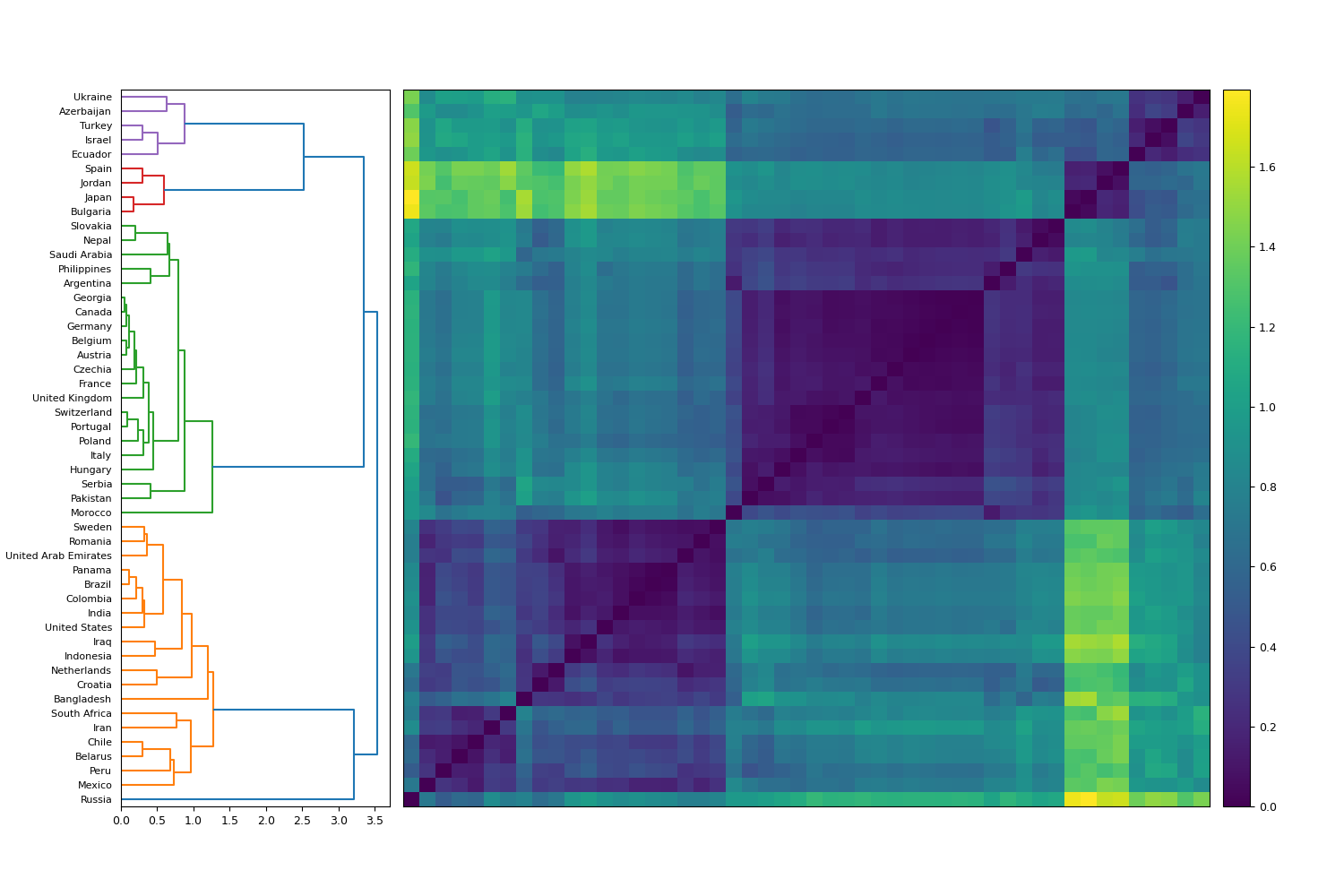}
        \caption{}
        \label{fig:MJ1/3}
    \end{subfigure}
    \begin{subfigure}[b]{\textwidth}
        \includegraphics[width=0.905\textwidth]{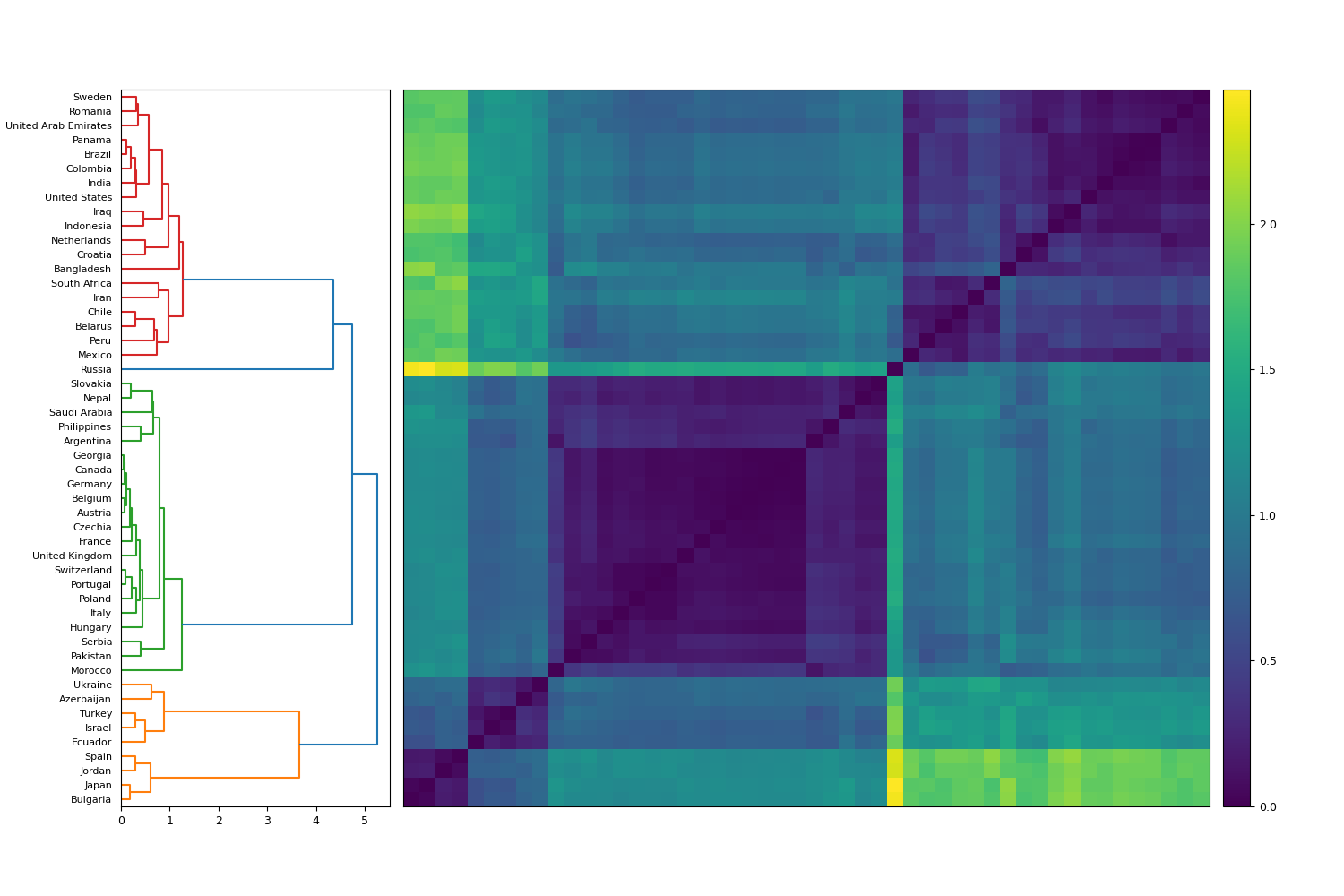}
        \caption{}
        \label{fig:MJ1/2}
    \end{subfigure}
    \caption{Hierarchical clustering on the turning point distance matrix $D^{TP}$, defined in Section \ref{sec:deathrates}, for (a) $\beta=1/3$, (b) $\beta=1/2$. This groups countries according to their similarity in undulating behaviour, measured by distances between turning point sets. Five characteristic classes are observed: Russia has two turning points; Brazil, India and the US have three; most European countries have four, with a strong subcluster of similarity observed including Austria, Belgium, and others. Two smaller classes are observed containing five and six turning points, respectively. The cluster structure in the two dendrograms is near identical, with a consolidation of the five- and six-turning point classes in (b). There, these classes are clearly observed as subclusters.}
    \label{fig:DeathRateDendrogramMJ1}
\end{figure*}

In Figure \ref{fig:DeathRateTrajectories}, we display rolling mortality rate and turning points for 12 countries: Brazil, India, Mexico, the United States (US), the Netherlands, Sweden, France, Germany, Italy, Russia, Ecuador and Bulgaria. These countries display highly heterogeneous behaviours, which are suitably captured in Figure \ref{fig:DeathRateDendrogramMJ1}. Figure \ref{fig:MJ1/3} reveals four clusters of similarity, and one outlier. Russia (\ref{fig:DeathRateRussia}) is the unique country with just two detected turning points. Several developing countries such as Brazil (\ref{fig:DeathRateBrazil}), India (\ref{fig:DeathRateIndia}) and Mexico (\ref{fig:DeathRateMexico}) as well as developed countries including the US (\ref{fig:DeathRateUS}), the Netherlands (\ref{fig:DeathRateNetherlands}) and Sweden (\ref{fig:DeathRateSweden}) have three turning points. France (\ref{fig:DeathRateFrance}), Germany (\ref{fig:DeathRateGermany}) and Italy (\ref{fig:DeathRateItaly}) have four turning points. Ecuador (\ref{fig:DeathRateEcuador}) and others have five, while Bulgaria (\ref{fig:DeathRateBulgaria}) and others have six. Figure \ref{fig:MJ1/2} gives a near-identical result, where the clusters pertaining to five and six turning points merge. However, examining the dendrogram closely, both are clearly visible as subclusters, and we are comfortable identifying five categories of trajectories. To demonstrate the robustness of our method, we record the cluster structure for a greater range of $\beta$ in Table \ref{tab:table_regularisation}.

Within the 4-turning point cluster, we see a dense subcluster of similarity containing Austria, Belgium, Canada, Czechia, France, Georgia, Germany, Hungary, Italy, Poland, Portugal, Switzerland and the United Kingdom (UK). All these countries experienced a peak in the mortality rate in April or May (corresponding to the previous 30 days) and a local minimum near the beginning of September (corresponding to the previous 30 days during August). This similarity can be seen by examining members of this cluster, France (\ref{fig:DeathRateFrance}), Germany (\ref{fig:DeathRateGermany}) and Italy (\ref{fig:DeathRateItaly}).

\begin{figure*}
    \centering
    \includegraphics[width=0.905\textwidth]{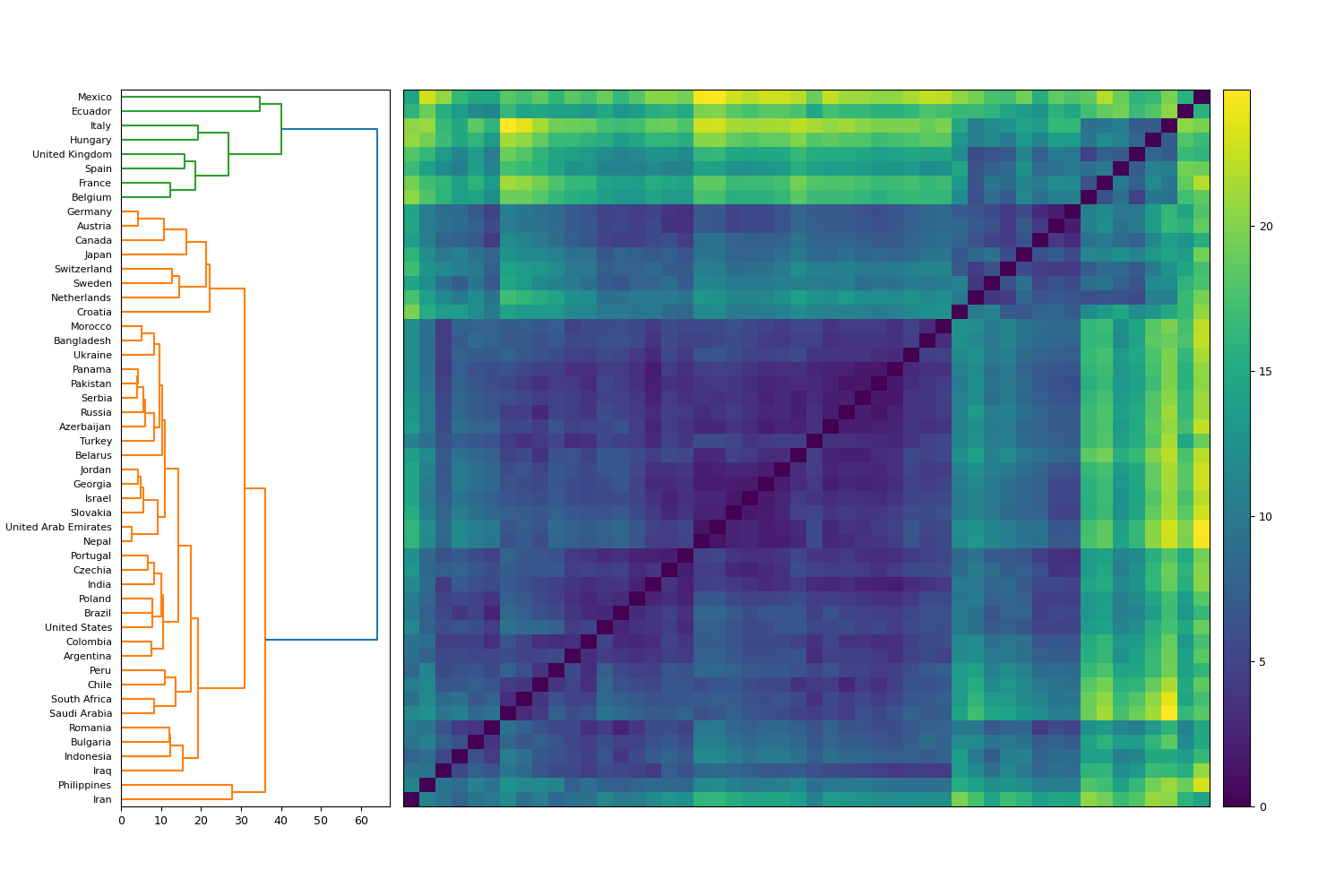}
    \caption{Hierarchical clustering on the $L^1$ distance matrix $D^{1}$, defined in Section \ref{sec:deathrates}. Mexico and Ecuador emerge as outliers, characterised by a consistently high mortality rate over the full period and the highest peaks in mortality of all, respectively. Belgium, France, Hungary, Spain and the UK are revealed as a secondary cluster, characterised by high mortality in April and May, rapidly decreasing from then.}
    \label{fig:DeathRateDendrogramTrajectory}
\end{figure*}

Turning to Figure \ref{fig:DeathRateDendrogramTrajectory}, several other insights concerning the mortality rate trajectories emerge. First, Mexico and Ecuador are identified as outliers in the collection of countries, with only slight similarity to each other. For Mexico (\ref{fig:DeathRateMexico}), this is due to a consistently high mortality rate over time, over 10\% for most of the period. Ecuador (\ref{fig:DeathRateEcuador}) is an outlier due to peaks in mortality over 30\%, higher than any other country. Belgium, France, Hungary, Spain, and the UK form their own smaller cluster characterised by high mortality rates (of around 20\%) in their first wave of COVID-19. Indeed, these countries experienced higher mortality in March-April than anywhere else in the world.

% Table
\begin{table}[ht]
\begin{center}
\begin{tabular}{ |p{1.5cm}||p{2cm}|p{2cm}|}
 \hline
 \multicolumn{3}{|c|}{Cluster robustness vs $\beta$} \\
 \hline
 $\beta$ & \# Clusters & Cluster sizes \\
 \hline
 1/5 & 4 & \{21,19,9,1\} \\
 1/4 & 4 & \{21,19,9,1\} \\
 1/3 & 5 & \{21,19,5,4,1\} \\
 1/2 & 4 & \{21,19,9,1\} \\
 1 & 4 & \{21,19,9,1\} \\
\hline
\end{tabular}
\caption{Number of clusters and cluster sizes for different values of the parameter $\beta$, used to define the semi-metric in Eq. (\ref{eq:newMJ}). While a different number of clusters is observed for $\beta=1/3$, subclusters with 5 and 4 elements are clearly visible in Figure \ref{fig:MJ1/3}.}
\label{tab:table_regularisation}
\end{center}
\end{table}

% Include comment on how 4th cluster splits into 5th (2 sub-clusters)

%-------------------SECTION III ----------
\section{Virulence matrix analysis}
\label{sec:virulencematrices}

\begin{figure}
    \centering
    \begin{subfigure}[b]{0.495\textwidth}
        \includegraphics[width=\textwidth]{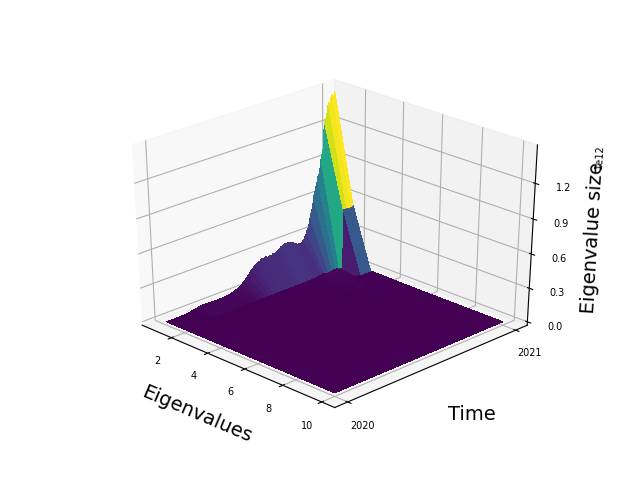}
        \caption{}
        \label{fig:VirulenceCases}
    \end{subfigure}
    \begin{subfigure}[b]{0.495\textwidth}
        \includegraphics[width=\textwidth]{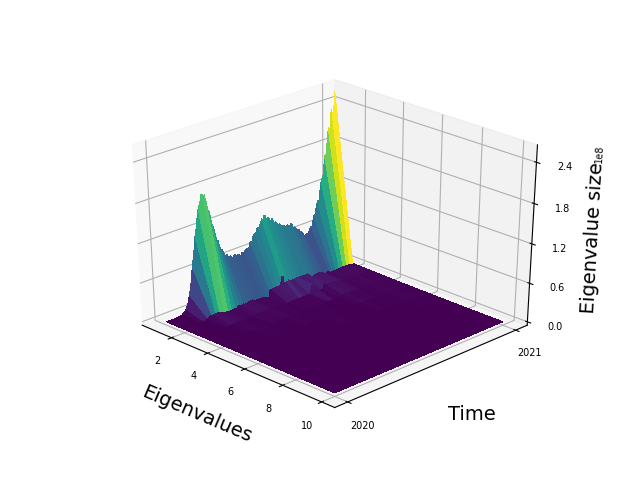}
        \caption{}
        \label{fig:VirulenceDeaths}
    \end{subfigure}
    \begin{subfigure}[b]{0.495\textwidth}
        \includegraphics[width=\textwidth]{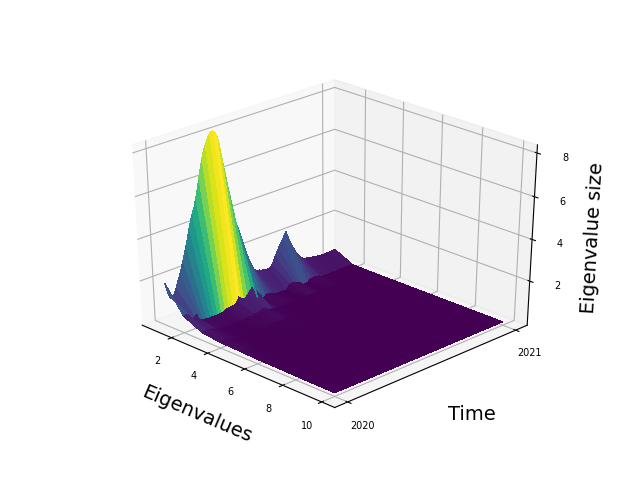}
        \caption{}
        \label{fig:VirulenceDeathRate}
    \end{subfigure}
    \caption{Time-varying eigenspectra (first ten eigenvalues) for the virulence matrices associated to (a) cases (b) deaths (c) mortality rate. The first eigenvalue demonstrates broad trends in the total size of the matrices, and shows (a) a large increase of cases towards the end of 2020, (b) two or three waves of significant deaths, (c) the highest mortality early on in the year. The second eigenvalue reveals more heterogeneity in case trajectories towards the end of the year, and mortality towards the beginning of the year.}
    \label{fig:VirulenceMatrices}
\end{figure}

In this section, we develop a new framework of time-varying analysis of 30-day rolling \emph{virulence matrices}, inspired by, but differing from, covariance matrices in finance \cite{Fenn2011}. Let $t=30,...,T$ be a particular time. We form vectors $\mathbf{x}_i(t)=(x_i(t-29),...,x_i(t))$, analogously for $\mathbf{y}_i$(t). These two vectors record the case and death counts over the past 30 days, with the subscript $i$ referring to the $i$th country, ordered alphabetically. We may also form $\mathbf{r}_i(t)=(r_i(t-29),...,r_i(t))$ for $t=59,...,T$, as the time series $r_i(t)$ only begin at $t=30$. Define (unscaled) inner products by
\begin{align}
    \langle \mathbf{x}_i(t),\mathbf{x}_j(t) \rangle = \sum_{s=t-29}^t x_i(s) x_j(s).
    %<\mathbf{x}_i(t),\mathbf{x}_j(t)>_n = \frac{\sum_{s=t-29}^t x_i(s) x_j(s)}{(x_i(t-29)^2+...+x_i(t)^2)^\frac{1}{2}(x_j(t-29)^2+...+x_j(t)^2)^\frac{1}{2}}
\end{align}
We then define $n \times n$ (unscaled) virulence matrices with respect to cases, deaths and mortality rates by the following $(i,j=1,...n)$:
\begin{align}
V_{ij}^{c}(t)&=  \langle \mathbf{x}_i(t),\mathbf{x}_j(t) \rangle, t=30,...,T;\\
%V_{ij}^{norm,case}(t)=   <\mathbf{x}_i(t),\mathbf{x}_j(t)>_n\\
V_{ij}^{d}(t)&=   \langle \mathbf{y}_i(t),\mathbf{y}_j(t) \rangle, t=30,...,T;\\
V_{ij}^{r}(t)&=   \langle \mathbf{r}_i(t),\mathbf{r}_j(t) \rangle, t=59,...,T.
%V_{ij}^{us,death}(t)=   <\mathbf{y}_i(t),\mathbf{y}_j(t)>_n
\end{align}
The subscript $ij$ refers to an inner product between the $i$th and $j$th countries, while the superscripts $c,d,r$ refer to case, death and mortality rate time series, respectively. Due to the summation procedure used to form these inner products, these virulence matrices implicitly average over 30 days' worth of case counts and can are thus robust against the noise present in day-to-day data. We could also analogously define normalised virulence matrices by using normalised inner products in place of the unscaled inner products above. These matrices are thus named because they provide a representation of the global spread of COVID-19 over the last 30 days and contain relationships between different countries' trajectories. The use of a standard covariance matrix here would not appropriately measure this prevalence: a country with a constant (but severe) number of cases for the past 30 days would yield a zero covariance with any other country. Each matrix $V(t)$ is a $n\times n$ symmetric real matrix, and thus is diagonalisable with all real eigenvalues. By the positivity of the inner product, each matrix satisfies a non-negativity condition $\mathbf{u}^TV\mathbf{u}\geq 0$ for $\mathbf{u}\in \mathbb{R}^n$, and so all eigenvalues are non-negative. We list and order the eigenvalues $\lambda_1\geq\lambda_2\geq...\geq\lambda_n\geq 0$. This produces a time-varying eigenspectrum, which we display in Figure \ref{fig:VirulenceMatrices} for the first ten eigenvalues. Moreover, for any such symmetric matrix, the greatest eigenvalue $\lambda_1$ holds particular significance. By the spectral theorem, $\lambda_1$ coincides with the \emph{operator norm} of the matrix \cite{RudinFA}, a measure of its total size. That is, 
\begin{align}
\lambda_1=||V||_{op}= \max_{\mathbf{u} \in \mathbb{R}^n - \{0\}} \frac{||V\mathbf{u}||}{||\mathbf{u}||}.
\end{align}
Subsequent eigenvalues also have a real-world interpretation. $\lambda_2=0$ if and only if the matrix $V$ is rank 1, which occurs if and only if all trajectories $\mathbf{x}_i$ (in the instance of the cases matrix) differ by a multiplicative constant. In general, a small value of $\lambda_2$ relative to $\lambda_1$ indicates substantial homogeneity in the trajectories.

In Figures \ref{fig:VirulenceCases}, \ref{fig:VirulenceDeaths} and \ref{fig:VirulenceDeathRate}, respectively, we display the time-varying eigenspectra for the virulence matrices associated to cases, deaths and mortality rates.  There are several interesting properties of these time-varying eigenspectra. The first eigenvalue $\lambda_1$ of Figure \ref{fig:VirulenceCases} demonstrates the general increase of new COVID-19 cases over the course of 2020. The sharp spike towards the end of the year demonstrates the rapid growth in cases in the final months of 2020. Figure \ref{fig:VirulenceDeaths} has two prominent peaks in its first eigenvalue, corresponding to the periods of March-April and November-December. These peaks highlight the natural history of COVID-19, where many countries suffered significant deaths during their first wave of the virus, enforced harsh restrictions resulting in fewer cases and deaths, and subsequently experienced further growth in cases and deaths upon such restrictions' easing. Finally, the first eigenvalue in Figure \ref{fig:VirulenceDeathRate} highlights an interesting trend in the mortality rate. There is a marked spike in March-April, followed by a significant reduction throughout the remainder of 2020. This shape in the first eigenvalue likely represents vulnerable people dying earlier and/or under-reporting of cases early in the year, contributing to a higher calculated mortality rate from reported cases and deaths.

The relationship between the first eigenvalue and subsequent eigenvalues is also of interest. Figure \ref{fig:VirulenceCases} shows the second eigenvalue $\lambda_2$ becoming quite significant for cases towards the end of 2020, when the total number of cases is larger than ever. This shows that the behaviour of new cases in late 2020 is more heterogeneous than the first wave, when all cases were rising quite uniformly throughout the world. Figures \ref{fig:VirulenceDeaths} and \ref{fig:VirulenceDeathRate} show a more moderate, but similar phenomenon concerning deaths and mortality rate at various stages of the year. The second eigenvalue in Figure \ref{fig:VirulenceDeaths} is slightly more pronounced in the second wave of the virus, displaying more heterogeneity in COVID-19 deaths later in the year. The second eigenvalue in Figure \ref{fig:VirulenceDeathRate} is more pronounced during the first wave of the virus - highlighting more heterogeneity during the first wave of the virus with respect to mortality. Indeed, Figure \ref{fig:DeathRateTrajectories} shows that European countries experienced substantial mortality in their first wave of COVID-19, which characterised them as anomalous in Figure \ref{fig:DeathRateDendrogramTrajectory}. This contributed to a meaningful heterogeneity of mortality rates across the world during the early stages of the year.

%VIRULENCE FINDINGS:
%cases - most severe and heterogeneous at end of year
%deaths - most severe in March/April and end of year, most hetero mid year
%mortality - most severe and heterogeneous in March/April

%-------------------SECTION IV-------

\section{Inconsistency analysis}
\label{sec:3wayCCA}
In this section, we describe how we measure the consistency between three attributes, and reveal anomalous countries in the process. To do so, we introduce a new method of comparing three distance matrices and apply this to distances between case and death time series, and human development indices (HDI). This generalises prior work studying anomalies between two attributes \cite{James2020_chaos}.

\subsection{Methodology}

Let $h_i$ be the HDI of each country. Calculated by the United Nations Development Programme \cite{UNHDR}, this index combines a country's life expectancy, educational standards and economic standard of living. Bounded between 0 and 1, the HDI $h_i$ reflects a substantially lower living standard the further $h$ moves from 1. To reflect this, we use a logarithmic distance between these indices that penalises movement away from 1 more than a linear distance:
\begin{align}
%D_{ij}^{c}&=||x_i - x_j||=\sum_{t=1}^T |x_i(t)-x_j(t)| \\
%D_{ij}^{d}&=||y_i - y_j||=\sum_{t=1}^T |y_i(t)-y_j(t)| \\
D_{ij}^{h}&=|\log h_i - \log h_j|, i,j=1,...,n.
\end{align}
As before, the subscript $i$ refers to the $i$th country, ordered alphabetically, the subscript $ij$ refers to a distance between the $i$th and $j$th country, while the superscript $h$ signals a distance relative to HDI. This forms a $n \times n$ distance matrix between countries' development indices. Given the exponential nature of the spread of the virus, we also use a logarithmic distance between the case and death time series. Some of these time series have negative counts due to retrospective adjustments in the data. In order to ensure non-negative counts, we first apply a Savitzky-Golay filter to produce smoothed case and death time series $\hat{x}_i(t)$ and $\hat{y}_i(t)$ respectively. Due to its moving average and polynomial smoothing, this eliminates almost all negatives, except when there are very few counts. We replace any non-positive count with a 1. Then, we may calculate a logarithmic $L^1$ distance as follows:
\begin{align}
D_{ij}^{c}&=||\log \hat{x}_i - \log \hat{x}_j||=\sum_{t=1}^T |\log \hat{x}_i(t)-\log \hat{x}_j(t)|; \\
D_{ij}^{d}&=||\log \hat{y}_i - \log \hat{y}_j||=\sum_{t=1}^T |\log \hat{y}_i(t)- \log \hat{y}_j(t)|.
%D_{ij}^{h}&=|\log h_i - \log h_j|, i,j=1,...,n
\end{align}
Above, the superscripts $c,d$ refer to distances relative to the case and death time series, respectively, between countries $i$ and $j$. Again, the summation across many days of data has the effect of smoothing out over the noise inherent in day-to-day variations of case counts. We use such a metric between case or death time series rather than a simple difference between the total yearly counts to distinguish between countries (and hence reveal potential anomalies) according to when the cases or deaths occurred. Thus, we have defined three $n \times n$ distance matrices between countries. Given a $n \times n$ distance matrix $D$, its corresponding \emph{affinity matrix} is defined as 
\begin{equation}
    A_{ij} = 1 - \frac{D_{ij}}{\max{\{D\}}}, i,j=1,...,n.
\end{equation}
All elements of these affinity matrices lie in $[0,1]$, so it is appropriate to compare them directly by taking their difference. Given a $n \times n$ matrix $C$, let $|C|$ be the matrix given by taking the absolute value of all elements, that is $|C|_{ij}=|C_{ij}|$. Then, define three $n \times n$ symmetric pairwise inconsistency matrices:
\begin{align}
        \text{INC}^{c,d} &= |A^{c}  - A^{d}|;\\  
         %\text{INC}^{c,d}_{ij} &= |A^{c}_{ij}  - A^{d}_{ij}|\\
         %\text{INC}^{c,h}_{ij} &= |A^{c}_{ij}  - A^{h}_{ij}|\\ 
         \text{INC}^{c,h} &= |A^{c}  - A^{h}|;\\
        \text{INC}^{d,h} &= |A^{d}  - A^{h}|;
%        \text{INC}^{d,h}_{ij} &= |A^{d}_{ij}  - A^{h}_{ij}|
\end{align}
%and analogously for $\text{INC}^{C,H}$ and $\text{INC}^{D,H}$. 
and a total inconsistency matrix
\begin{align}
    \text{INC}^{c,d,h}=\text{INC}^{c,d}+\text{INC}^{c,h}+\text{INC}^{d,h}.
\end{align}
Above and below, a superscript $c,d$ refers to inconsistency between cases and deaths, while $c,h$ refers to an inconsistency between cases and HDI, and similarly for $d,h$. A superscript $c,d,h$ refers to an inconsistency between all three attributes. Next, we can define pairwise anomaly scores by 
\begin{align}
    %a_i^{C,D}=\sum_{j=1}^n |\text{INC}_{ij}^{C,D}|\\
     a_i^{c,d}&=\sum_{j=1}^n \text{INC}_{ij}^{c,d};\\
  a_i^{c,h}&=\sum_{j=1}^n \text{INC}_{ij}^{c,h};\\
     a_i^{d,h}&=\sum_{j=1}^n \text{INC}_{ij}^{d,h}.
\end{align}

\begin{figure}
    \centering
        \includegraphics[width=\textwidth]{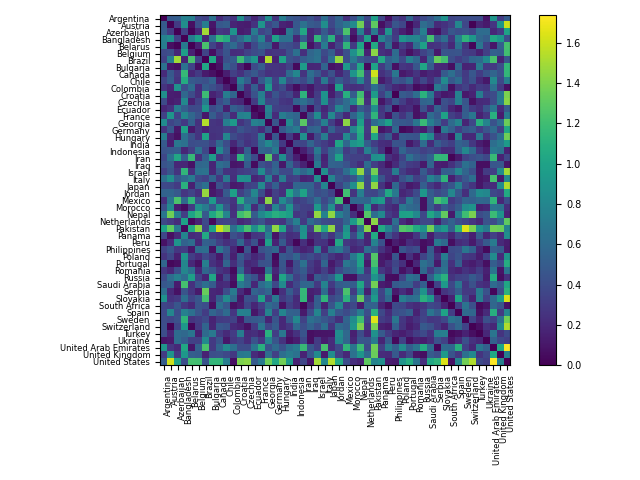}
        \label{fig:CCA_Consistency}
    \caption{Total anomaly matrix $\text{INC}^{c,d,h}$, as defined in Section \ref{sec:3wayCCA}. Lighter entries indicate higher values of the matrix, and hence more inconsistency between the attributes under consideration: cases, deaths and HDI. The US and Pakistan can be seen to have substantial inconsistency with many other countries.}
    \label{fig:3wayCCA}
\end{figure}

%and analogously for $a_i^{C,H}$ and $a_i^{D,H}$. 
For each country, we record an \emph{anomaly vector} $\mathbf{a}_i= (a_i^{c,d},a_i^{c,h}, a_i^{d,h})$ and a \emph{total anomaly score} given by $a_i^{c,d,h}=a_i^{c,d}+a_i^{c,h}+a_i^{d,h}$. We can also define a \emph{weighted anomaly score} to reduce bias in one set of anomaly scores being systematically larger than another. Let $M^{c,d}=\max_i \{a_i^{c,d}\}$, analogously for $M^{c,h}$ and $M^{d,h}$. Let the weighted anomaly score be $\tilde{a}_i^{c,d,h}=a_i^{c,d}/M^{c,d}+a_i^{c,h}/M^{c,h}+a_i^{d,h}/M^{d,h}$. This aims to record a neutral contribution from each anomaly score. In Tables \ref{tab:3waycca1} and \ref{tab:3waycca2}, we record the anomaly vectors, total anomaly score and weighted anomaly score for all 50 countries under consideration. In Figure \ref{fig:3wayCCA}, we plot the total consistency matrix $\text{INC}^{c,d,h}$, where anomalous countries can easily be seen due to larger entries in their respective rows and columns. An analogous weighted consistency matrix can also be defined, which is broadly similar to the one shown.

\begin{remark}
In this brief aside, we explore the edge cases of maximal consistency and maximal inconsistency, and interpret their meaning. Consider a single entry $\text{INC}^{c,d}_{ij}=|A^{c}_{ij}  - A^{d}_{ij}|$. As both $A^{c}_{ij}, A^{d}_{ij} \in [0,1]$, the inconsistency entry $\text{INC}^{c,d}_{ij}$ has greatest possible value to 1. It attains that value when $A^{c}_{ij}=1$ and $A^{d}_{ij}=0$, or vice versa. This equations can be reinterpreted as $D^{c}_{ij}=0$ and $D^{d}_{ij}= \max D^d$, respectively. That is, greatest inconsistency occurs when countries $i$ and $j$ have equal case counts, but the greatest difference in death counts among any pair of countries, or vice versa. The exact same statement applies for greatest inconsistency between cases and HDI or deaths and HDI.

On the other hand, greatest possible consistency across the entire matrix would mean $A^{c}_{ij}  - A^{d}_{ij}=0$, for all $i,j$. Rearranging this yields $\frac{D^{c}_{ij}}{\max D^{c}} = \frac{D^{d}_{ij}}{\max D^{d}}$, for all $i,j$. That is, the distance matrices $D^c$ and $D^d$ differ up to a single scalar. One example where this can occur is if there are constants $a$ and $\tau$ such that $y_i(t)=ax_i(t+\tau)$, for all $i=1,...,n, t=1,...,T$. Then this relationship passes to the smoothed counts by linearity, and so $\sum_t |\log \hat{y}_i(t)- \log \hat{y}_j(t)| = \sum_t |\log a\hat{x}_i(t+\tau)- \log a\hat{x}_j(t+\tau)| =  \sum_t |\log \hat{x}_i(t)- \log \hat{x}_j(t)|$. 
%As some countries may have zero cases and deaths, the only practically possible value for $b$ is $b=0$. 
That is, maximal consistency would occur if every country has an identical progression of cases to deaths, up to a multiplicative constant $a$ and a time-offset $\tau$.
\end{remark}

\subsection{Results}

The total inconsistency matrix and all computed anomaly scores yield several insights. First, the three most anomalous countries with respect to the weighted anomaly score are Pakistan, the US and the United Arab Emirates (UAE). A near-identical result applies if we use the unscaled total anomaly score, with Pakistan, the US, Nepal and then the UAE exhibiting the largest unscaled scores. For the US and Pakistan, the highest contribution to the total or weighted anomaly score comes from their high pairwise anomaly scores $a^{c,h}$ and $a^{d,h}$, which are the two highest of any country. Interestingly, these high scores have differing explanations. The US is highly inconsistent between cases (and analogously deaths) and HDI due to its much higher case and death counts than other countries of similar HDI. Pakistan is classified as inconsistent due to an extreme HDI, the lowest of any country under consideration, but a case and death time series that are similar to many others. Thus, due to a lower HDI than other countries with similar case and death counts, it is registered as inconsistent. We remark that high anomaly scores do not necessarily indicate a straightforward anomalous quotient between cases or HDI, for example. Instead, a high anomaly score reflects inconsistency in relationships with other countries.

On the other hand, the UAE has a high weighted and total anomaly score due to its value of $a^{c,d}$, which is the highest of any country. Indeed, the UAE experienced the lowest mortality rate across 2020 of any country under consideration. The country with the second-highest value of $a^{c,d}$ is Mexico. This is anomalous for the opposite reason: a consistently high progression from cases to deaths, as first noted in Figure \ref{fig:DeathRateMexico}.

\begin{table}
\begin{tabular}{|p{2.5cm}|p{1.3cm}|p{1.3cm}|p{1.3cm}|p{1.3cm}|p{1.3cm}|}
 \hline
 \multicolumn{6}{|c|}{Country anomaly scores relative to cases, deaths and HDI (1)} \\
 \hline
 Country & $a^{c,d}$ &  $a^{c,h}$ &$a^{d,h}$ & $a^{c,d,h}$ & $\tilde{a}^{c,d,h}$ \\
 \hline
Argentina & 3.23 & 8.96 & 10.59 & 22.78 & 1.98 \\
Austria & 3.43 & 8.73 & 10.45 & 22.61 & 1.20 \\
Azerbaijan & 3.87 & 7.47 & 9.72 & 21.06 & 1.15 \\
  Bangladesh & 2.47 & 14.69 & 14.93 & 32.10 & 1.58 \\
  Belarus & 4.41 & 7.86 & 9.91 & 22.19 & 1.23 \\
  Belgium & 3.88 & 8.27 & 8.52 & 20.67 & 1.13 \\
  Brazil & 4.04 & 12.29 & 15.31 & 31.64 & 1.64 \\
  Bulgaria & 2.26 & 8.84 & 8.31 & 19.41 & 0.99 \\
  Canada & 2.87 & 9.15 & 9.21 & 21.23 & 1.11 \\
  Chile  & 3.00 & 9.18 & 10.85 & 23.04 & 1.20 \\
  Colombia  & 3.55 & 6.28 & 6.48 & 16.30 & 0.92 \\
  Croatia  & 2.91 & 10.53 & 11.04 & 24.48 & 1.26 \\
  Czechia  & 4.06 & 8.23 & 9.86 & 22.15 & 1.21 \\
  Ecuador  & 3.75 & 7.52 & 6.96 & 18.23 & 1.02 \\
  France  & 2.98 & 9.77 & 10.70 & 23.45 & 1.21 \\
  Georgia  & 4.86 & 13.33 & 11.92 & 30.11 & 1.61 \\
  Germany  & 3.33 & 9.83 & 8.36 & 20.62 & 1.10 \\
  Hungary  & 3.52 & 10.74 & 8.94 & 23.21 & 1.23 \\
  India  & 2.75 & 9.67 & 9.72 & 22.14 & 1.14 \\
  Indonesia  & 3.72 & 8.37 & 7.47 & 19.56 & 1.07 \\
  Iran  & 5.25 & 8.77 & 11.68 & 25.69 & 1.43 \\
  Iraq  & 3.42 & 7.35 & 5.88 & 16.65 & 0.93 \\
  Israel  & 4.98 & 9.74 & 11.15 & 25.86 & 1.43 \\
  Italy  & 4.03 & 9.27 & 11.67 & 24.96 & 1.34 \\
  Japan  & 2.79 & 9.75 & 10.34 & 22.88 & 1.18 \\
  Jordan  & 4.66 & 11.06 & 10.94 & 26.66 & 1.44 \\
  Mexico  & 9.73 & 6.82 & 13.06 & 29.60 & 1.85 \\
  Morocco  & 3.97 & 10.34 & 9.67 & 24.00 & 1.29 \\
  Nepal  & 4.53 & 16.84 & 17.78 & 39.15 & 2.00 \\
  Netherlands  & 3.97 & 8.41 & 8.70 & 21.08 & 1.16 \\
  Pakistan  & 2.43 & 22.73 & 21.65 & 46.80 & 2.24 \\
  Panama  & 3.34 & 7.17 & 7.92 & 18.42 & 1.01 \\
  Peru  & 3.06 & 6.48 & 6.98 & 16.52 & 0.90 \\
  Philippines  & 2.51 & 7.49 & 7.29 & 17.29 & 0.91 \\
  Poland  & 2.68 & 7.69 & 8.58 & 18.94 & 0.99 \\
  Portugal  & 3.11 & 7.42 & 8.10 & 18.63 & 1.00 \\
  Romania  & 3.18 & 7.24 & 7.68 & 18.10 & 0.98 \\
\hline
\end{tabular}
\caption{Anomaly vectors, total anomaly scores and weighted anomaly scores, as defined in Section \ref{sec:3wayCCA}, for the first 37 countries under consideration. Pairwise anomaly scores quantify the inconsistency in measurements between two quantities, while the total and weighted anomaly scores incorporate all three attributes. The weighted anomaly score is chosen to more appropriately weight the contributions from the three pairwise scores.}
\label{tab:3waycca1}
\end{table}

\begin{table}
\begin{tabular}{|p{2.5cm}|p{1.3cm}|p{1.3cm}|p{1.3cm}|p{1.3cm}|p{1.3cm}|}
 \hline
 \multicolumn{6}{|c|}{Country anomaly scores relative to cases, deaths and HDI (2)} \\
 \hline
 Country & $a^{c,d}$ &  $a^{c,h}$ &$a^{d,h}$ & $a^{c,d,h}$ & $\tilde{a}^{c,d,h}$ \\
 \hline
  Russia  & 2.62 & 10.62 & 10.75 & 23.99 & 1.21 \\
  Saudi Arabia  & 3.10 & 10.77 & 10.54 & 24.41 & 1.26 \\
  Serbia  & 3.57 & 7.98 & 9.80 & 21.36 & 1.15 \\
  Slovakia  & 3.34 & 12.86 & 13.13 & 29.33 & 1.50 \\
  South Africa  & 3.16 & 7.01 & 5.72 & 15.89 & 0.88 \\
  Spain  & 4.05 & 9.82 & 11.27 & 25.14 & 1.35 \\
  Sweden  & 4.23 & 9.57 & 9.36 & 23.16 & 1.26 \\
  Switzerland  & 3.14 & 8.52 & 9.73 & 21.38 & 1.13 \\
  Turkey  & 2.50 & 7.81 & 7.78 & 18.08 & 0.95 \\
  Ukraine  & 2.78 & 6.86 & 6.45 & 16.08 & 0.87 \\
  UAE  & 10.29 & 8.78 & 13.56 & 32.63 & 2.01 \\
  UK  & 3.78 & 9.87 & 10.80 & 24.44 & 1.30 \\
  US  & 3.18 & 18.46 & 19.81 & 41.45 & 2.04 \\
\hline
\end{tabular}
\caption{Anomaly vectors, total anomaly scores and weighted anomaly scores, as defined in Section \ref{sec:3wayCCA}, for the remaining 13 countries under consideration. Pairwise anomaly scores quantify the inconsistency in measurements between two quantities, while the total and weighted anomaly scores incorporate all three attributes. The weighted anomaly score is chosen to more appropriately weight the contributions from the three pairwise scores.}
\label{tab:3waycca2}
\end{table}

\section{Discussion}
\label{sec:conclusion}

%\textcolor{blue}{We believe that this is the first paper to quantitatively combine these features and identify respective anomalies.} - PUT IN CONCLUSION

%Intro to conclusion
In this paper, we analyse the natural history of COVID-19 across 50 countries over 2020. We observe significant structural similarity between certain countries as well as heterogeneity across the world with respect to COVID-19 prevalence and mortality, and identify anomalous countries therein. Such insights cannot be gained with conventional techniques, such as a comparison of reproductive ratios across countries. Our analysis consistently considers the changing dynamics with time.

In Section \ref{sec:deathrates}, we analyse mortality rate trajectories for 50 countries. By modifying a recently introduced turning point algorithm and introducing a new semi-metric between turning point sets, we assign these time series into five characteristic classes according to their differing trajectories. Russia is identified as an outlier - its mortality rate rose consistently until July and never dropped substantially enough to register a subsequent trough in our algorithmic framework. It is unique in this sense among the 50 countries, possessing a consistently stable mortality rate after its first peak. 19 countries exhibit three turning points, including Brazil, India and the US, indicating a substantial reduction in mortality from a first peak. 21 countries exhibit four turning points, indicating a second wave in which mortality has increased once again. In particular, a strong subcluster contains most Western European countries: Austria, Belgium, Czechia, France (\ref{fig:DeathRateFrance}), Germany (\ref{fig:DeathRateGermany}), Hungary, Italy (\ref{fig:DeathRateItaly}), Portugal, Switzerland, and the UK. These all share highly similar mortality trajectories, with a first peak in April-May, a trough around September, and another peak at the end of the year.

There are three wealthy western European countries that do not fit into this cluster. Both the Netherlands and Sweden, displayed in Figures \ref{fig:DeathRateNetherlands} and \ref{fig:DeathRateSweden} respectively, do not register a second peak in mortality. Indeed, these countries both kept their mortality low towards the end of the year, while France, Germany and Italy experienced an increase. Prior research has noted that the Netherlands reduced its mortality rate substantially in its second wave of COVID-19 \cite{jamescovideu}, while Sweden changed its COVID-19 response substantially relative to the first half of the year \cite{Sweden}. Spain registers six turning points primarily due to highly irregular reporting, featuring negative counts and large numbers of cases and deaths consolidated and reported on single sporadic days.

A smaller number of countries exhibited more turning points: five with 5 turning points and four with 6. We observe that the majority of developed countries exhibit 3 or 4 turning points, as visible in Figure \ref{fig:DeathRateDendrogramMJ1}, while the outlier countries (with 2,5 or 6 turning points) were mostly developing countries. This reflects more regular (and less undulating) behaviour in the mortality rate trajectories and has two explanations. First, more developed countries may have implemented more consistent testing, which could have caused less fluctuations in the reported mortality rate. Secondly, more developed countries may have more healthcare resources to improve their treatment of COVID-19 and thereby reduce and stabilise the mortality rate over time.

As a whole, the most significant finding from Section \ref{sec:deathrates} is the identification of five categories of mortality trajectories, attained from both the turning point algorithm and the use of clustering our new semi-metric between sets. This reveals close similarity among mortality trajectories when considered as varying functions over time, and carries more weight than an overall comparison of mortality obtained by just dividing the number of deaths observed throughout 2020 by the number of cases. Our results are robust with respect to the variation of parameters.
%As mortality changes over time, it is essential to analyse over time, and observe structural similarity between countries in its dynamics. Our results are robust with respect to the variation of parameters.

In Section \ref{sec:virulencematrices}, we introduce a new class of virulence matrices for cases, deaths and mortality rates and analyse their eigenspectra. The first eigenvalue $\lambda_1$ provides a measure of the total scale of the matrices and summarises worldwide trends in prevalence and mortality throughout 2020. 
%Viewed together, these three trends provide several insights. 
Figure \ref{fig:VirulenceCases} reflects a substantial surge in cases towards the end of the year, Figure \ref{fig:VirulenceDeaths} shows multiple waves of deaths of comparable magnitude, while Figure \ref{fig:VirulenceDeathRate} shows an early peak that dominates the rest of the period. The second eigenvalue $\lambda_2$ provides a measure of the heterogeneity among the studied time series. Figure \ref{fig:VirulenceCases} exhibits a considerable rise in heterogeneity towards the end of the year, during a time in which new cases trajectories across different countries were substantial but quite non-uniform. In Figure \ref{fig:VirulenceDeaths}, we see a much greater value of $\lambda_2$ during the second wave of deaths, in which $\lambda_1$ is in fact lower than the first wave. The much milder drop off between $\lambda_1$ and $\lambda_2$ indicates the greatest heterogeneity with respect to deaths during this period in the middle of the year. Figure \ref{fig:VirulenceDeathRate} similarly reveals substantial heterogeneity in mortality rates during the earlier part of the year.

When viewed in conjunction, these three figures provide several insights into the natural history of the disease throughout 2020. Case counts generally increased in global severity throughout the year, while death counts constituted a much clearer pattern of multiple waves. The mortality rate trajectory (\ref{fig:VirulenceDeathRate}) can explain this - in March and April, the progression from reported cases to deaths was much more severe throughout Europe, causing substantial deaths despite fewer cases than late 2020. During the middle of the year, the heterogeneity in death counts was at its highest. Indeed, the months of June to August featured relatively few new cases in Europe \cite{NPRJune}, while Brazil \cite{Reutersbrazil}, India and other developing countries experienced substantial growth in cases \cite{global_health_2020}. Towards the end of the year, the pandemic once again impacted the entire world, with more counts observed than ever before. During this time, mortality was low, but cases were so high that deaths became the highest they have ever been. Heterogeneity in case trajectories also increased substantially, with COVID-19 trajectories differing substantially between different countries, many increasing, some decreasing, but most with high total counts. One could more closely examine heterogeneity by considering normalised virulence matrices obtained from normalised inner products, as explained in Section \ref{sec:virulencematrices}.

This analysis provides a new means of identifying periods of maximal severity and heterogeneity in case, death and mortality trajectories across the world. %, considering the world's counts as a multivariate system. 
The temporal dimension is critical in such analyses as both severity and heterogeneity change over time. 
%The main finding of this section is the identification of specific time intervals of greatest severity and heterogeneity across three different time series.
Specifically, cases are most severe and heterogeneous at the end of the year; deaths are most severe in March/April and year-end, but most heterogeneous in the middle of the year; mortality is most severe and heterogeneous in March/April.

%VIRULENCE FINDINGS:
%cases - most severe and heterogeneous at end of year
%deaths - most severe in March/April and end of year, most hetero mid year
%mortality - most severe and heterogeneous in March/April

In Section \ref{sec:3wayCCA}, we study the consistency between cases, deaths and HDI for all 50 countries under consideration. We believe that this is the first method proposed to study (in)consistencies among a collection of time series for up to three measures. We propose two measures of anomaly across these three quantities: a total and weighted anomaly score (that more appropriately combines the contributions of the three pairwise anomaly components). The three most anomalous countries with respect to the weighted score are Pakistan, the US and the UAE. Closer inspection of the pairwise anomaly components in Tables \ref{tab:3waycca1} and \ref{tab:3waycca2} can reveal which quantities most contribute to a country's total or weighted score. For the UAE, this is the high anomaly score between cases and deaths, caused by the lowest progression from cases to deaths among our collection of countries. For the US, both anomaly scores $a^{c,h}$ and $a^{d,h}$ contribute highly; these reflect the fact that the US has considerably more cases and deaths than other countries of similar HDI. For Pakistan, the same two anomaly scores $a^{c,h}$ and $a^{d,h}$ are the largest of any country, but for the opposite reason: its HDI is substantially lower than any country with a similar case and death time series.

The full collection of anomaly scores can also reveal broad trends regarding consistency between the three measures. In Tables \ref{tab:3waycca1} and \ref{tab:3waycca2}, we see that the two pairwise anomaly scores relative to HDI are systematically greater than the pairwise score between case and death counts. Indeed, we have $a_i^{c,d} < a_i^{d,h}$ for every single country and $a_i^{c,d} < a_i^{c,h}$ for every country except Mexico (which has the second-highest case-death anomaly score after the UAE due to its consistently and anomalously high mortality). These patterns reveal systematically more consistency between case and death counts than between case or death counts and HDI. Qualitatively, this reveals there is little relationship between a country's HDI and its case or death counts. In addition, a closer examination reveals that $a_i^{c,h} < a_i^{d,h}$ for 34 out of the 50 countries, 2/3 of the collection. Thus, to a lesser extent, there is greater consistency between case counts and HDI than there is between death counts and HDI. This is a surprising finding - one would naively expect more consistency between a lower HDI and higher deaths due to poorer healthcare quality resulting in a greater progression of cases to deaths, regardless of the number of cases.

The originality of Section \ref{sec:3wayCCA} is two-fold: first, a new mathematical method for identifying inconsistencies across three attributes; and second, as the first analysis of cases, deaths and HDI of different countries simultaneously, again taking temporal dynamics into account. The main findings are the identification of specific anomalous countries, including Mexico and the UAE between cases and deaths, and the US and Pakistan between cases (or deaths) and HDI.

Several limitations and opportunities for future research exist in this inconsistency framework. First, the results could also be repeated for case and death time series as a proportion of each country's population. Alternative metrics between cases and deaths could be used, such as a simple difference between the total yearly counts, without the temporal component provided by the $L^1$ metric. 
%We use such a metric between case or death time series rather than a simple difference between the total yearly counts to distinguish between countries (and hence reveal potential anomalies) according to when the cases or deaths occurred.
A closer analysis of the relationship between the varying sizes of the anomaly scores could quantitatively characterise the differing consistency between three quantities as a whole. One limitation in this analysis framework is that anomalies are measured purely by their relative deviation from the rest of the collection, and direction (positive or negative) is ignored. A closer inspection is necessary to determine the nature of the anomaly. However, this could be seen as a benefit of the methodology as well, as it is flexible in the detection of different sorts of inconsistent behaviour. 
%ADDITIONAL HERE
Further research could also incorporate several different attributes other than HDI, such as countries' age demographics, size, and population density.

%A first one is the different number of diagnostic tests (per capita) performed in different countries what would affect substantially the time series. Also many other factors playing a role are country demographics (e.g. population above 70), country size, fraction of population living in (large) cities, etc.

More broadly speaking, any analysis of reported cases and deaths due to COVID-19 will have limitations. First, the reported counts of COVID-19 may have been under-reported \cite{underreporting} throughout the pandemic. Not only did early cases spread throughout Europe and the US before testing programs had been established, but testing protocols were far from uniform across the year and between countries. Indeed, several countries changed their testing protocols on various occasions, including within the same wave \cite{Francechange,Pullano2020,Antigenchange}. 
%Within Italy, for example, different regions operated according to different protocols, testing only symptomatic patients or more broadly. \cite{DiBari2020}
Even deaths may have been under-reported, with substantial differences observed between excess mortality and reported COVID-19 deaths \cite{italydeaths}. Nonetheless, our analysis of reported case and death counts may reveal structural similarity and anomalies, help governments in their decision-making, and motivate further research that examines other data attributes in more involved studies.

%The same limitations as previously stated apply, due to countries changing their testing protocols even within the second wave, and varying between countries. \cite{Antigenchange}

\section{Conclusion}
\label{sec:finalconclusion}

%Final wrapup
Overall, this paper introduces new methods for analysing COVID-19 prevalence and mortality on a country-by-country and worldwide basis and chronicles the natural history of COVID-19 during 2020. On a global scale, we reveal broad trends in case and death counts as well as mortality trajectories, which present a coherent picture of the changing impacts of COVID-19 over time. On a country-by-country basis, we reveal both heterogeneity and structural similarity with respect to mortality over time and study consistency between COVID-19 prevalence and human development, revealing specific anomalous countries. Moreover, the framework presented in this paper could be applied broadly to various epidemiological or economic crises. The consistent theme in our analysis, and motivation for it, is to always seek structure and associated anomalies in case, death and mortality time series, with an essential consideration of changing dynamics with time.

The primary strength of this analysis is that our findings are difficult to detect with existing methods. For example, the use of SIR models and their extensions, together with an analysis of the reproductive ratio $R_0$, may be fit independently for each country and create predictions, but they are not suitable to detecting structure in all the world's case, death and mortality trajectories at once. Such methods would not reveal the five classes of trajectories we find, nor would they identify the periods of the greatest heterogeneity in prevalence or mortality, nor identify anomalous countries with respect to our chosen data attributes. Measurements such as $R_0$ are more useful for early analysis of the transmissibility of the virus; we aim to find structure while comparing the plights of different countries over time.

% The consistent theme in our analysis, and motivation for it, is to examine the continual evolution of structure and associated anomalies over time. Given the complex nature of COVID-19, some countries have high cases and low deaths, low deaths and high cases, etc. - so it's important to examine more than cases/deaths or mortality rates in isolation. Accordingly, we introduce new methodologies to examine the time-varying structure and anomalous progression of cases, deaths and mortality time series.

%What governments can take out of it
As 2021 begins, the world remains severely affected by COVID-19. Though vaccination distribution is underway in many countries, the analysis of trends in cases, deaths and mortality remains of substantial relevance to governments. The identification of structural similarity in mortality rate trajectories between European states may inspire additional cooperation \cite{Priesemann2021} and coordination of their strategic response to the pandemic. Our methods highlight countries that have responded particularly well or poorly, and our analysis highlights points in time where cases, deaths and mortality rates changed substantially for candidate countries. Finally, we reveal global changes in the relationship between cases, deaths and mortality rates over time. Such changes should inform governments regarding their response to the pandemic. This will be particularly crucial in the coming months, as various vaccines are administered over the world.

%----- END OF PAPER------------------

\section*{Acknowledgements}
Many thanks to Kerry Chen for helpful comments and edits.

\section*{Data availability}
Daily COVID-19 case and death counts and human development index data can be accessed at "Our World in Data" \cite{worldindata2020}.

\section*{Funding sources}
This research did not receive any specific grant from funding agencies in the public, commercial, or not-for-profit sectors.

\appendix

\section{Turning point methodology}
\label{appendix:turningpoint}

In this section, we provide more details for the identification of turning points of a mortality rate time series $r(t)$. First, some smoothing is necessary due to irregularities in the data set, and discrepancies between different data sources. The Savitzky-Golay filter ameliorates these issues by combining polynomial smoothing with a moving average computation, and yields a smoothed time series $\hat{r}(t) \in \mathbb{R}_{\geq 0}.$ Subsequently, we perform a two-step process to select and then refine a non-empty set $P$ of local maxima (peaks) and $T$ of local minima (troughs).

Following \cite{james2020covidusa}, we apply a two-step algorithm to the smoothed time series $\hat{r}(t)$. The first step produces an alternating sequence of troughs and peaks. The second step refines this sequence according to chosen conditions and parameters. The most important conditions to identify a peak or trough, respectively, in the first step, are the following:
\begin{align}
\label{baddefnpeak}
\hat{r}(t_0)&=\max\{\hat{r}(t): \max(1,t_0 - l) \leq t \leq \min(t_0 + l,T)\},\\
\label{baddefntrough}\hat{r}(t_0)&=\min\{\hat{r}(t): \max(1,t_0 - l) \leq t \leq \min(t_0 + l,T)\},
\end{align}
where $l$ is a parameter to be chosen. Due to the smoothing of the Savitzky-Golay filter, noise in day-to-day counts will change the local maxima and minima of the smoothed time series minimally, and will not affect either the number of total turning points nor the distances between different turning point sets. Following \cite{james2020covidusa}, we select $l=17$, which accounts for the 14-day incubation period of the virus \cite{incubation2020} and less testing on weekends. Defining peaks and troughs according to this definition alone has several flaws, such as the potential for two consecutive peaks.

Instead, we implement an inductive procedure to choose an alternating sequence of peaks and troughs. Suppose $t_0$ is the last determined peak. We search in the period $t>t_0$ for the first of two cases: if we find a time $t_1>t_0$ that satisfies (\ref{baddefntrough}) as well as a non-triviality condition $\hat{r}(t_1)<\hat{r}(t_0)$, we add $t_1$ to the set of troughs and proceed from there. If we find a time $t_1>t_0$ that satisfies (\ref{baddefnpeak}) and  $\hat{r}(t_0)\geq \hat{r}(t_1)$, we ignore this lower peak as redundant; if we find a time $t_1>t_0$ that satisfies (\ref{baddefnpeak}) and  $\hat{r}(t_1) > \hat{r}(t_0)$, we remove the peak $t_0$,  replace it with $t_1$ and continue from $t_1$. A similar process applies from a trough at $t_0$. 

At this point, the time series is assigned an alternating sequence of troughs and peaks. However, some turning points are immaterial and should be excluded. The second step is a flexible approach introduced in \cite{james2020covidusa} for this purpose. In this paper, we introduce new conditions within this framework. First, we use the same peak ratio procedure: let $t_1<t_3$ be two peaks, necessarily separated by a trough. We select a parameter $\delta=0.2$, and if the \emph{peak ratio}, defined as $\frac{\hat{r}(t_3)}{\hat{r}(t_1)}<\delta$, we remove the peak $t_3$. If two consecutive troughs $t_2,t_4$ remain, we remove $t_2$ if $\hat{r}(t_2)>\hat{r}(t_4)$, otherwise remove $t_4$. That is, if the second peak has size less than $\delta$ of the first peak, we remove it.

Finally, let $t_1,t_2$ be adjacent turning points (one a trough, one a peak). We choose a parameter $\epsilon=\log(2)$;  if
\begin{align}
    |\log \hat{r}(t_2) - \log \hat{r}(t_1)|<\epsilon,
\end{align}
that is, the values of the turning point differ by less than a factor of 2, we remove $t_2$ from our sets of peaks and troughs. If $t_2$ is not the final turning point, we also remove $t_1$. This is a different condition from previous work - whereas \cite{james2020covidusa} considers the average change with time between turning points of new case trajectories, we consider only the absolute change between turning points in mortality rate. Indeed, there is no need to consider how much time has passed when determining whether mortality has increased or decreased by a sufficient amount, in our implementation a factor of 2, to warrant a turning point being included. 

%mortality rate is bounded between 0 and 1, unlike new case counts, and we do not consider how much time is taken for mortality rate to change by a certain amount (in this case a factor of 2)

%It is more suited to this application, where mortality rate trajectories are not expected to increase or decrease bounded by a particular logarithmic rate; instead, changes in mortality may be quite drastic with time.

\bibliographystyle{elsarticle-num-names}
\bibliography{1References.bib}
%\end{nolinenumbers}
\end{document}